\documentclass[12pt]{iopart}

\usepackage{graphicx}
\usepackage{dcolumn}
\usepackage{bm}
\usepackage{xcolor}
\usepackage{verbatim}

\begin{document}
\title[Electromagnetic field diagnostics with an electron plasma]{\textit{In situ} electromagnetic field diagnostics with an 
electron plasma in a Penning-Malmberg trap}


\author{C. Amole$^1$\footnote{Present Address: Department of Physics, Engineering Physics and Astronomy, Queen's University, Kingston ON, K7L 3N6, Canada},
 M. D. Ashkezari$^2$,
 M. Baquero-Ruiz$^3$,
 W. Bertsche$^{4,5}$,
 E. Butler$^6$\footnote{Present address: Centre for Cold Matter,
Imperial College, London SW7 2BW, United Kingdom},
        A. Capra$^1$,
        C. L. Cesar$^7$,
        M. Charlton$^8$,
        A. Deller$^8$,
        N. Evetts$^9$,
        S. Eriksson$^8$,
        J. Fajans$^{3,10}$,
        T. Friesen$^{11}$,
        M. C. Fujiwara$^{12}$,
        D. R. Gill$^{12}$,
        A. Gutierrez$^9$,
        J. S. Hangst$^{13}$,
        W. N. Hardy$^9$,
        M. E. Hayden$^2$,
        C. A. Isaac$^8$,
        S. Jonsell$^{14}$,
        L. Kurchaninov$^{12}$,
        A. Little$^3$,
        N. Madsen$^8$,
        J. T. K. McKenna$^{15}$,
        S. Menary$^{1}$,
        S.C. Napoli$^{14}$,
        K. Olchanski$^{12}$,
        A. Olin$^{12}$,
        P. Pusa$^{15}$,
        C. \O{}. Rasmussen$^{13}$,
        F. Robicheaux$^{16}$\footnote{Present Address: Department of Physics, Purdue University, West Lafayette, IN 47907, USA},
        E. Sarid$^{17}$,
        D. M. Silveira$^7$,
        C. So$^3$,
        S. Stracka$^{12}$\footnote{Present Address: Scuola Normale Superiore, 56126 Pisa, Italy},
        T. Tharp$^3$,
        R. I. Thompson$^{11}$,
        D. P. van der Werf$^8$,
        J. S. Wurtele$^3$}
\address{$^1$Department of Physics and Astronomy, York University, Toronto ON, M3J 1P3, Canada}
\address{$^2$Department of Physics, Simon Fraser University, Burnaby BC, V5A 1S6, Canada}
\address{$^3$Department of Physics, University of California, Berkeley, CA 94720-7300, USA}
\address{$^4$School of Physics and Astronomy, University of Manchester, Manchester M13 9PL, UK}
\address{$^5$The Cockcroft Institute, Warrington WA4 4AD, UK}
\address{$^6$Physics Department, CERN, CH-1211 Geneva 23, Switzerland}
\address{$^7$Instituto de F\'{i}sica, Universidade Federal do Rio de Janeiro, Rio de Janeiro 21941-972, Brazil}
\address{$^8$Department of Physics, College of Science, Swansea University, Swansea SA2 8PP, UK}
\address{$^9$Department of Physics and Astronomy, University of British Columbia, Vancouver BC, V6T 1Z4, Canada}
\address{$^{10}$Lawrence Berkeley National Laboratory, Berkeley, CA 94720, USA}
\address{$^{11}$Department of Physics and Astronomy, University of Calgary, Calgary AB, T2N 1N4, Canada}
\address{$^{12}$TRIUMF, 4004 Wesbrook Mall, Vancouver BC, V6T 2A3, Canada}
\address{$^{13}$Department of Physics and Astronomy, Aarhus University, DK-8000 Aarhus C, Denmark}
\address{$^{14}$Department of Physics, Stockholm University, SE-10691 Stockholm, Sweden}
\address{$^{15}$Department of Physics, University of Liverpool, Liverpool L69 7ZE, UK}
\address{$^{16}$Department of Physics, Auburn University, Auburn, AL 36849-5311, USA}
\address{$^{17}$Department of Physics, NRCN-Nuclear Research Center Negev, Beer Sheva, 84190, Israel}

\newpage

\begin{abstract}

We demonstrate a novel detection method for the cyclotron resonance frequency of an electron plasma in a Penning-Malmberg trap. 
With this technique, the electron plasma is used as an \textit{in situ} diagnostic tool for measurement of the 
static magnetic field and the microwave electric field in the trap. The cyclotron motion of the electron plasma is excited by
microwave radiation and the temperature change of the plasma is measured non-destructively by monitoring the plasma's 
quadrupole mode frequency. The spatially-resolved
microwave electric field strength can be inferred from the plasma temperature change and the magnetic field 
is found through the cyclotron resonance frequency. These measurements were used extensively in the recently reported 
demonstration of resonant quantum interactions with antihydrogen. 

\end{abstract}

\maketitle

\section{Introduction}

Cyclotron frequency measurements of single particles and sparse clouds in Penning traps are commonly used in 
high precision ion mass measurements~\cite{Marshall1998,Bergstrom2002,Dilling2006}
and in measurements of the proton~\cite{DiSciacca2012} and electron~\cite{Odom2006} magnetic moments. In the
plasma regime, the cyclotron resonances of electron~\cite{Gould1991} and ion~\cite{Sarid1995} plasmas have also been studied 
extensively. 
Cyclotron resonances of ions (or electrons in a low magnetic field) typically occur at radio frequencies and are relatively 
easy to detect. Electron cyclotron frequencies, however, are often at high microwave frequencies and must be detected
using alternative methods. In the single particle regime, microwave cyclotron frequencies are measured using methods that couple 
the cyclotron 
and axial motions~\cite{Dyck1981,Odom2006}, producing detectable shifts in the axial bounce frequency. Here we outline and
demonstrate a novel detection method of the cyclotron resonance of an electron plasma at microwave frequencies. The key feature of 
our method is the use of the quadrupole, or breathing, mode oscillation of the electron plasma to detect excitation of the 
cyclotron motion. We focus on the use of this technique as a tool for in-situ characterization of the magnetic field and a 
microwave field in a Penning-Malmberg trap. While we demonstrate the technique for microwave frequencies, this method can in 
principle be applied to cyclotron resonances in the radio frequency range. 

Initially this work was motivated by the need for an \textit{in situ} measurement of the static magnetic field in the ALPHA 
(Antihydrogen 
Laser PHysics Apparatus) experiment at CERN (European Organization for Nuclear Research)~\cite{Friesen2013}. 
ALPHA and ATRAP (Antihydrogen TRAP, another CERN-based experiment) synthesize neutral antihydrogen atoms from their charged 
constituents, held in Penning-Malmberg 
traps. Low-energy 
antihydrogen atoms are then confined in magnetic potential wells (Ioffe-Pritchard type~\cite{Pritchard1983} magnetic minimum 
atom traps), thereby 
eliminating interactions with (and annihilations on) the walls of the surrounding apparatus~\cite{Andresen2010,Gabrielse2012}. 
Accurate \textit{in situ} 
determinations of these trapping fields will play an important role in future precision antihydrogen spectroscopy experiments. 
In this work, electron plasmas are confined along the common axis of the Penning-Malmberg and Ioffe-Pritchard traps and are
used to probe the magnetic trapping fields in the vicinity of the minimum magnetic field. This is precisely the region of the 
trap that is of spectroscopic interest; energy intervals between hyperfine levels of the antihydrogen atom are field dependent, 
leading to the appearance of sharp extrema in transition frequencies as atoms pass through the field minimum~\cite{Ashkezari2012}. 
The methods 
presented here were used extensively in the recent demonstration of the first resonant electromagnetic interaction with 
antihydrogen~\cite{Amole2012}.

We also discuss the use of an electron plasma as a microwave electric field probe. Knowledge of 
the microwave field is crucial for any microwave experiment in such a trap. Gaps between electrodes, changes in electrode radius, 
and reflections create an environment that supports a complex set of standing and travelling wave 
modes. The resulting microwave electric fields can vary drastically as a function of position and frequency and accurately 
simulating the mode structure presents a largely intractable problem. Using the magnitude of the cyclotron heating by a microwave 
field, however, we can estimate the amplitude of the co-rotating component of the electric field.
Furthermore, by employing techniques analogous to magnetic resonance imaging, we can create a map of the 
co-rotating microwave electric field amplitude along the cylindrical Penning-Malmberg trap axis. 

\section{Method}

For the remainder of this work we focus on the implementation of the techniques in a cylindrical Penning-Malmberg trap. In 
principle, the techniques could be adapted for implementation in a hyperbolic electrode Penning trap.  We operate
under the assumption that the cyclotron frequency of the electron plasma is equivalent to the single particle cyclotron frequency, 
$f_{\mathrm{c}} = qB/2 \pi m$, where $q$ is the electron charge, $B$, is the amplitude of the magnetic field, and $m$ is 
the electron mass. In general, a non-neutral plasma will have a set of cyclotron modes that are shifted away from the single 
particle frequency, an issue that we discuss in section~\ref{shifts}. In the measurements presented here, however, such frequency 
shifts are below the achieved resolution. 

When the cyclotron motion of electrons in a plasma is driven by a pulsed microwave field, the absorbed energy is 
redistributed through collisions,
resulting in an increased plasma temperature. We measure this temperature change by non-destructively probing the plasma's 
quadrupole mode frequency. 
The quadrupole mode oscillation of a non-neutral plasma is just one of a set of electrostatic plasma modes~\cite{Dubin1991}. The 
frequencies of these modes are set by the plasma density, temperature, and aspect ratio 
$\alpha = L/2r$, where $L$ is the plasma length (major axis) and $r$ is the radius (semi-minor axis). 
For a plasma confined in a perfect quadratic potential produced by distant electrodes, these mode frequencies can be calculated 
analytically in the cold-fluid limit~\cite{Dubin1991} and have been used experimentally to make non-destructive measurements of 
plasma parameters~\cite{Tinkle1994,Amoretti2003,Speck2007}. The quadrupole mode holds particular interest here because 
the frequency is shifted with increasing temperature above the cold fluid limit. An approximate treatment 
of non-zero temperatures has been proposed~\cite{Dubin1993} and shown to agree well with experiment~\cite{Amoretti2003,Tinkle1995}.
For a change in plasma temperature by $\Delta T$, the corresponding change in frequency is given by

\begin{equation} ({f_2}^{'})^2 - (f_2)^2 = 5\left(3 -
  \frac{\alpha^2}{2}\frac{f_p^2}{(f_2^c)^2}\frac{\partial^2
    g(\alpha)}{\partial \alpha^2} \right)\frac{k_B\Delta T}{m\pi^2L^2}, \label{quad_shift}\end{equation}

\noindent where $k_B$ is the Boltzmann constant, and $f_2$ and ${f_2}^{'}$ are the quadrupole frequencies before and
after the heating pulse, respectively. The quadrupole frequency in the cold fluid limit is given by $f_2^{\mathrm{c}}$ and 
$g(\alpha) = 2Q_1[\alpha(\alpha^2 - 1)^{-1/2}]/(\alpha^2 - 1)$, where $Q_1$ is the first order Legendre function of the second kind. 
The plasma frequency $f_\mathrm{p}$ is given by $f_{\mathrm{p}}=(2\pi)^{-1}(nq^2/m\epsilon_0)^{1/2}$, where $n$ is the plasma number
density, and $\epsilon_0$ is the permittivity of free space. The temperature dependence of $f_2$ can be used to realize a 
non-destructive plasma temperature diagnostic. We typically work in a regime where $\Delta f_2/f_2 \ll 1$, with 
$\Delta f_2 = f_2^{'} - f_2$, so we make
the approximation $({f_2}^{'})^2 - (f_2)^2 \approx 2f_2 \Delta f_2$. The quadrupole frequency increase is therefore
expected to be linear with respect to the plasma temperature and given by
\begin{equation} \Delta f_2 \approx \beta \Delta T, \label{DW}\end{equation}
where $\beta$ is
\begin{equation} \beta = \frac{5}{2 f_2} \left(3 -
  \frac{\alpha^2}{2}\frac{f_p^2}{(f_2^c)^2}\frac{\partial^2
    g(\alpha)}{\partial \alpha^2} \right)\frac{k_B}{m\pi^2L^2}. \label{beta}\end{equation}
Calculation of $\beta$ from (\ref{beta}) assumes a plasma confined in a perfect harmonic potential and imperfections will shift
$\beta$ in an unknown manner. In the experiments that follow, the electrode structure has not been optimized to produce a perfect
harmonic potential. We instead experimentally determine $\beta$ as well as confirm the validity of (\ref{DW}) using an 
independent, destructive, temperature diagnostic (see section~\ref{quad_calib} for details). It is also important to note that the 
quadrupole mode frequency is used only to measure relative changes in plasma temperature and we do not attempt to infer absolute 
temperatures from the mode frequency.

The cyclotron resonance frequency is determined by monitoring the quadrupole mode frequency while a series of excitation 
pulses are applied at frequencies that scan through the cyclotron resonance. Each excitation pulse will cause a
jump in the quadrupole mode frequency, the amplitude of which should be maximized when the excitation frequency matches the 
cyclotron frequency.
Between each excitation pulse, the plasma cools back to its equilibrium
temperature via emission of cyclotron radiation. Because the quadrupole mode diagnostic is non-destructive, we can map out a full 
cyclotron lineshape using a single electron plasma.

In section~\ref{cycfreq} we demonstrate this technique in two different magnetic field profiles. The first is the standard 
uniform solenoidal field of a Penning-Malmberg trap. In this field, we expect a peak in the plasma heating at the single particle 
cyclotron resonance with a linewidth set by the temperature of the plasma
and any inhomogeneities in the magnetic field. We can also apply the cyclotron frequency measurements to measure the minimum 
magnetic field of a magnetic neutral atom trap such as that used for the trapping of antihydrogen 
in the ALPHA experiment. 

These methods can also be applied in a microwave electrometry mode by 
using the magnitude of plasma heating at the cyclotron frequency as a measure of the amplitude of the 
microwave electric field.  We can estimate 
the amplitude of the co-rotating component of the electric field by treating the electron 
plasma as a collection of single particles precessing around the magnetic field at the single
particle cyclotron frequency. Working from the single particle equations of motion for an electron:
\begin{equation} m\frac{d\mathbf{v}}{dt} = q\mathbf{E} + q\mathbf{v}\times\mathbf{B}, \end{equation}
we can find the average change in the transverse kinetic energy when a collection of electrons undergoing cyclotron motion 
is exposed to a near resonant transverse oscillating electric field. To
simplify the equations, we define $\omega = qB/m$ 
and decompose the electric field into components that co-/counter-rotate with respect to the
cyclotron motion; that is, $E_\pm(t) = E_x(t) \pm iE_y(t)$. We can then write $v_{\pm} = v_x(t) \pm iv_y(t)$ and
the single particle equations of motion become
\begin{equation} \frac{dv_{\pm}(t)}{dt} = \frac{q}{m} E_{\pm}(t) \mp i\omega v_{\pm}(t) \label{diff}.  \end{equation}
Assuming the heating pulses, and consequently $E_\pm(t)$, are non-zero only for a time short 
compared to damping and collisional timescales, the solution to (\ref{diff}) is
\begin{equation} v_\pm(t) = \left[v_\pm(t_0)e^{\pm i\omega t_0} + \frac{q}{m}\int_{-\infty}^t e^{\pm i\omega t'}E_\pm(t')dt'\right]e^{\mp i\omega t}, \end{equation}
where $E_\pm(t) = 0$ for $t < t_0$.
The change in average transverse kinetic energy, $\langle\mathrm{KE_{\perp}}\rangle = m\langle v_+v_- \rangle/2$, caused by
a pulse of microwaves is therefore
\begin{equation} \Delta\langle\mathrm{KE_{\perp}}\rangle = 
\frac{q^2}{2m}\left|\int_{-\infty}^{\infty}E_+(t')e^{i\omega t'}dt'\right|^2 \label{short_soln},\end{equation}
where $E_+ = E_x(t) + iE_y(t)$ is the co-rotating component of the microwave electric field. 
Following the microwave pulse, collisions 
redistribute the kinetic energy among the three degrees of freedom resulting in a temperature change of 
\begin{equation} \Delta T = \frac{2}{3k_B}\Delta \langle\mathrm{KE_{\perp}}\rangle. 
\label{DT}\end{equation}

By measuring the temperature increase due to a pulse of microwave radiation, via the quadrupole mode frequency increase, 
the magnitude of
the co-rotating microwave electric field can be calculated using (\ref{short_soln}) and (\ref{DT}). Microwave fields at 
different
frequencies can be probed by adjusting the magnetic field to set the cyclotron resonance to the desired frequency.
For convenience we will abbreviate the co-rotating microwave electric field as `CMEF' for the remainder of this work.

The structure of the CMEF along the trap axis can also be probed in this manner. If allowed by the trap 
construction, the electron plasma can be moved to different axial positions, providing a map of the CMEF 
strength at a resolution set by the plasma length. For a given plasma length, this resolution can be improved by making a magnetic
resonance imaging style scan of the plasma. A linear magnetic field gradient is applied across the length of the plasma, creating
a position dependant cyclotron frequency. Microwaves injected at a given frequency will only be resonant with a narrow slice of 
the plasma. The resulting plasma heating depends on the local CMEF over the narrow slice and the number of particles in resonance.
If the static magnetic field is changed, without changing the gradient, a different slice of 
plasma will be moved into resonance. In a uniform microwave electric field, this would amount to a one-dimensional projection
image of the plasma, 
with a plasma heating proportional to the number of particles in resonance at each step. In the case of a highly variable electric
field over the plasma length and an approximately uniform density spheroidal plasma, we can extract a map of the CMEF
strength over the plasma length.

\section{Apparatus}

\begin{figure}
\centering \includegraphics[width=0.75\columnwidth]{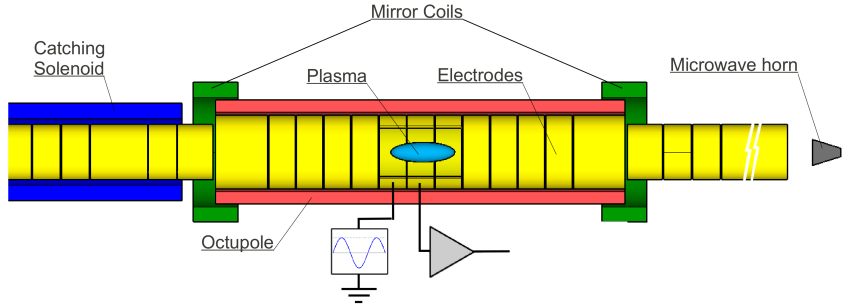}
\caption{Sketch of the core of the ALPHA apparatus. A 1 T solenoid (not pictured) surrounds the components shown here with
the exception of the microwave horn located 1.3 m from the centre of the trap. The quadrupole mode excitation signal is applied 
to an electrode at one end of the plasma and the subsequent ringing of the plasma is picked up from the centre electrode. 
The radial extent of the plasma has been exaggerated here for illustration purposes.}
\label{setup}
\end{figure}

The measurements presented here were performed by the ALPHA antihydrogen experiment~\cite{Amole2014} located in the Antiproton 
Decelerator facility
at CERN. The ALPHA Penning-Malmberg trap consists of 35 cylindrical electrodes whose axis is aligned with the axis of a 1 T  
superconducting solenoid. Both DC potentials and oscillating fields up to several tens of megahertz in frequency can be applied 
to the electrodes. The measurements were made in a region of the trap with an electrode wall radius of 22.5 mm.
The electrodes are thermally connected to a liquid helium bath and cool to approximately 7.5 K. 
Surrounding the trap electrodes are three superconducting magnets that form the magnetic minimum neutral atom trap 
(see figure~\ref{setup}). A 
three dimensional magnetic minimum is created by two mirror coils, which produce an axially increasing field around the centre,
and an octupole winding, creating a radially increasing field~\cite{Bertsche2006}. A smaller superconducting solenoid 
surrounds a portion of the Penning trap and is used in the capture of antiprotons from the Antiproton 
Decelerator. With the exception of section~\ref{Standingmap} this solenoid is not energized for any of the measurements presented 
here. 

Electrons are emitted by an electron gun positioned on the Penning-Malmberg trap axis by a moveable vacuum manipulator, which also
includes a micro-channel plate (MCP) and phosphor screen detector (collectively referred to as the `MCP detector' for convenience) 
and a microwave horn. Using the MCP detector, the plasma's integrated radial density profile can be measured 
destructively~\cite{Andresen2009}. The number of electrons in a trapped plasma can be measured by releasing the particles onto a 
Faraday cup and measuring the deposited charge. From the particle number, radial profile, and knowledge of the confining potentials,
 the full three-dimensional density distribution can be calculated numerically~\cite{Prasad1979}. The plasma radial distribution, 
and therefore the aspect ratio and the parameter $\beta$ (see (\ref{DW}),(\ref{beta})), can be manipulated by 
applying a `rotating-wall' electric field using segmented electrodes~\cite{Huang1997}. Plasma temperatures can also be measured 
using the MCP detector~\cite{Eggleston1992}. If the confining potential is slowly (with respect to the axial bounce frequency of
approximately 15 MHz) reduced, the highest energy particles will 
escape the well first and their charge will be registered by the MCP detector. Typically, the confining well is reduced to zero 
over 20 ms, ensuring that particles of a given energy have time to escape before the confining potential changes significantly. 
Assuming the plasma is in local thermal equilibrium along the magnetic field lines, the velocity distribution of the first escaping 
particles will follow the tail of a Maxwell-Boltzmann distribution~\cite{Eggleston1992}. The plasma temperature 
can therefore be determined by an exponential fit to the number of particles released as a function of well depth. 

Microwaves at frequencies between 26 and 30 GHz are generated by an Agilent 8257D synthesizer and are
carried by coaxial cable down one of two potential paths: a high power path with a
4 W amplifier for resonant experiments with antihydrogen, and a low power path (no amplification) for the 
electron cyclotron resonance diagnostics discussed here. These two paths merge just before entering the trap
vacuum via WR28 waveguide through a hermetically sealed quartz window. Finally, an internal length of waveguide brings the 
microwaves to a microwave horn that is aligned with the Penning trap axis. 

We measure the quadrupole mode frequency by first exciting the mode with a Gaussian modulated radio-frequency (RF) 
pulse applied to an 
electrode at one end of the plasma (see figure \ref{setup}). The subsequent ring down of the plasma ($Q \approx 1000$) is picked up 
on the 
central electrode. The response signal is amplified, passed through a broad band-pass filter, then digitized. The 
quadrupole mode frequency is extracted from the digital signal using a Fast Fourier Transform (FFT) and a peak-finding routine. 
The drive pulses are typically 0.3 - 1.0 V, 1 $\mu$s in duration, and thus have a spectral width of approximately 
1 MHz. We apply 5 pulses, each separated by 100 ms  and average the 5 response signals before performing the FFT. In this 
configuration the quadrupole mode frequency is probed every 1.2 s. 

All experiments in this paper utilize electron plasmas loaded in a roughly harmonic potential produced by five electrodes
in the centre of the Penning traps. At 1 T the electron cyclotron frequency is approximately 28 GHz. 
The plasmas typically have a radius of 1 mm and are 20 - 40 mm in length, overlapping three electrodes. Plasma loads of 
$3\times10^6$ to $4\times10^7$ electrons were studied. The lower limit is set by our ability to distinguish the quadrupole mode 
signal from background noise. These plasmas have densities between $5\times10^{13}$ and $5\times10^{14}$ m$^{-3}$ and typical 
temperatures of $\sim$150~K. At these densities and temperatures, collisions will bring the cyclotron motion into equilibrium
with the motion parallel to the magnetic field at a rate of roughly $10^{5}$~s$^{-1}$~\cite{Glinsky1992}. 
The quadrupole mode frequency of these plasmas is typically 24 - 28 MHz. 

\section{Quadrupole mode calibration}\label{quad_calib}

\begin{figure}
\centering \includegraphics[width=0.5\columnwidth]{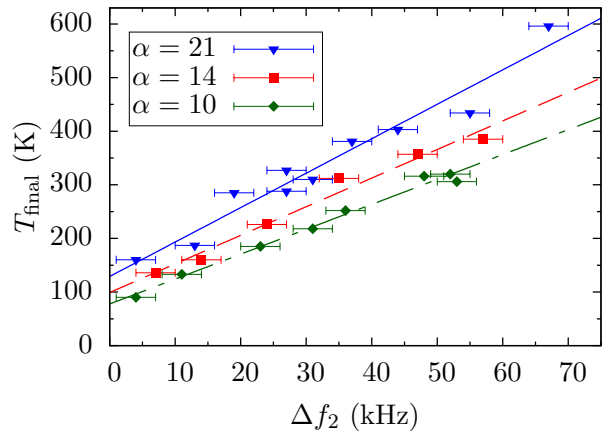}
\caption{Final plasma temperature plotted against the quadrupole frequency increase for three different plasma aspect ratios. 
The temperature measurement uncertainties are roughly 5 K and not visible on this plot. From a linear fit to the data 
$\beta^{-1}$ is 
calculated to be $6.4\pm0.5$~K/kHz for $\alpha = 21$, $5.3\pm0.4$~K/kHz for $\alpha=14$, and $4.6\pm0.2$ K/kHz for $\alpha=10$.}
\label{calib}
\end{figure}

Before using the quadrupole mode frequency shift to measure the cyclotron frequency or estimate the microwave electric field, 
the linearity of (\ref{DW}) and the value of $\beta$ were determined experimentally. 
This was accomplished by continuously monitoring the quadrupole mode frequency of an electron plasma, while an RF noise drive, 
applied to a nearby electrode, heats the plasma. After the plasma reaches a new equilibrium, the temperature is destructively 
measured using
the MCP detector. For different RF drive amplitudes, the final plasma temperatures and the corresponding quadrupole 
frequency shifts were determined. Figure~\ref{calib} shows the measured calibrations for three plasmas with the same number of 
electrons ($2\times10^7$) but different aspect ratios. The aspect ratios are determined from the numerically calculated
self-consistent density distributions based on measured particle numbers and radial profiles. 

The frequency shift is seen to be linear with the change in temperature, as predicted. 
For the majority of the measurements that follow, we use a plasma consisting of $1.2\times10^7$ electrons, $\alpha = 16$, $L = 26$
mm, a base temperature of $\sim$150 K, and a measured $\Delta T$ vs $\Delta f_2$ calibration of $\beta^{-1} = 3.7\pm0.3$~K/kHz. 
Quadrupole mode frequency changes can be measured to around 5 kHz, enabling us to resolve temperature changes of roughly 
20 K or greater with these plasma parameters.

\section{Cyclotron frequency measurements}\label{cycfreq}

The cyclotron frequency of an electron plasma is measured by repeatedly probing the quadrupole mode frequency while a series of 
microwave pulses are applied. Each microwave pulse is 4 $\mu$s long and at a different frequency, spanning a range that includes 
the cyclotron resonance. The resulting increase in the quadrupole mode frequency is then measured for each pulse. Between the
pulses, the electrons will radiatively cool in the 1 T field with a characteristic cooling time of roughly 4 s. To ensure the plasma
returns to thermal equilibrium, each pulse is separated by 15 - 35 s. 

\subsection{Uniform field}\label{Uniform}

\begin{figure}
\includegraphics[]{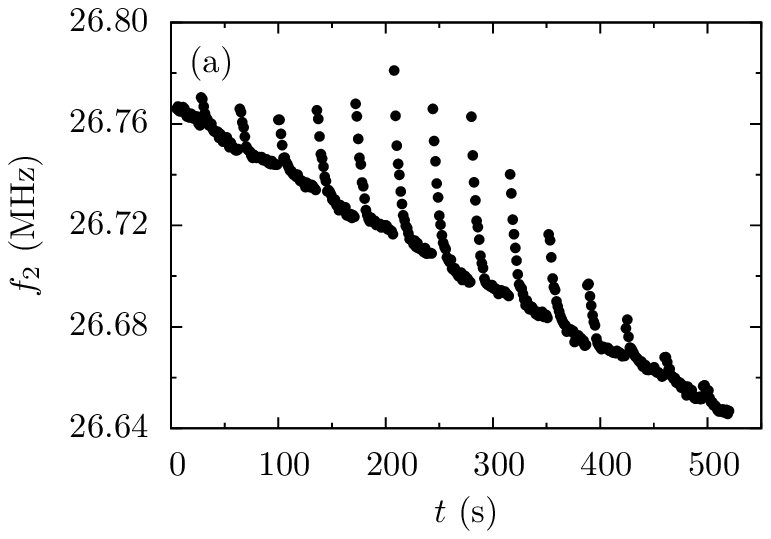}
\hspace{0.75cm}
\includegraphics[]{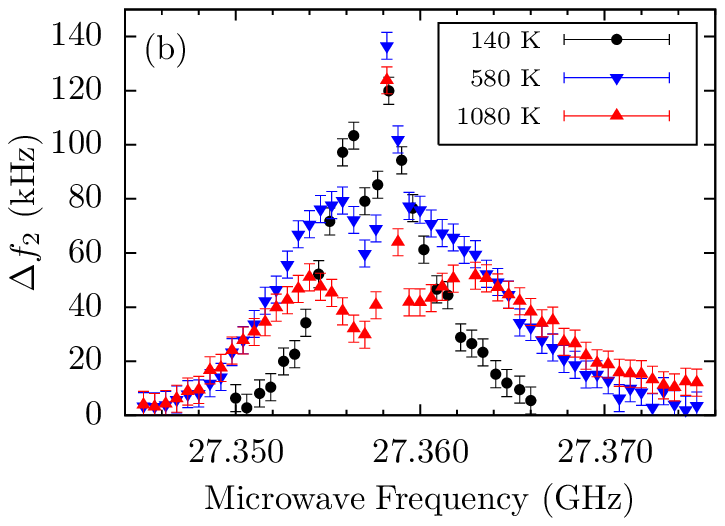}
\caption{(a) Real-time readout of the quadrupole mode frequency during a cyclotron resonance
scan of an electron plasma. The sudden jumps in frequency are due to 4~$\mu$s microwave pulses 
near the cyclotron resonance frequency. The slowly decreasing baseline quadrupole 
frequency is likely due to the slow expansion of the plasma. (b) The measured cyclotron lineshapes at different plasma temperatures.}
\label{uniformfield}
\end{figure}

We first examine the case of a nominally uniform solenoid field at 1 T. 
A real-time readout of the quadrupole frequency during a cyclotron frequency measurement can be seen in 
figure~\ref{uniformfield}(a). 
The lineshape is constructed by plotting the quadrupole mode frequency shifts ($\Delta f_2$) against the microwave frequency
(see figure \ref{uniformfield}(b)). The observed lineshape (without additional heating) is roughly Gaussian with a dip 
near the peak. 
Increasing the plasma temperature (via RF heating) broadens the overall lineshape
but a very strong narrow peak emerges with broad side lobe-like features (figure~\ref{uniformfield}(b)). This peak
does not appear to broaden as plasma temperature increases. 
At the end of the lineshape measurement, while still heating and probing the quadrupole mode frequency, we move the MCP 
detector into place to measure the plasma temperature. 

Similar datasets collected using a different cylindrical Penning-Malmberg trap show the same general features: a large roughly 
central peak with broad side lobe-like features. The side lobes and the relative height of the central peak change significantly 
at different cyclotron frequencies. Interpretation of these lineshapes is complicated by the strong frequency and position 
dependence of the microwave field. The narrow central peak is particularly surprising as its full-width-at-half-maximum (FWHM) is 
on the order of 0.5 - 1 MHz. For comparison, if the microwaves are treated as a plane-wave propagating down the trap axis a 
Doppler width of $\sim$10 MHz would be expected for an electron cloud at 150 K. The narrow width of the central peak may be
due to a subset of the electrons if they are confined within a constant phase region of a standing wave 
structure~\cite{Kleppner1962} in the trap or an effect of the fact the wavelength of the microwaves ($\sim$1 cm) is comparable to
the radius ($\sim$ 0.1 cm) and length ($\sim$4 cm) of the plasma~\cite{Dicke1953}. 

\begin{figure}
\begin{center}
\includegraphics{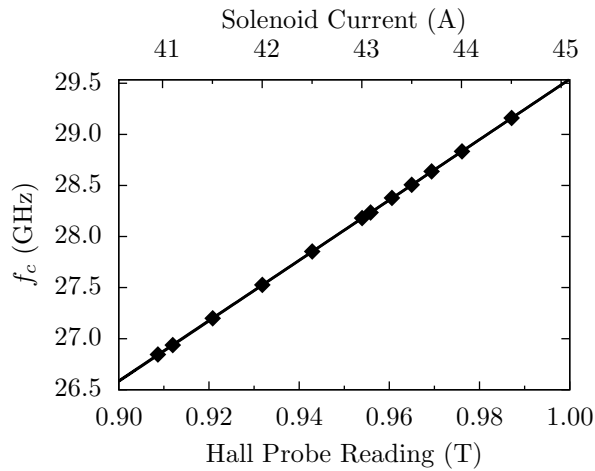}
\end{center}
\caption{Peak heating frequency as a function of the solenoid current and resulting magnetic field as measured by a 
Hall probe. The Hall probe is located within the bore of the Penning trap solenoid.}
\label{lakeshorecalib}
\end{figure}

While we do not have a complete understanding of the observed lineshapes, the position of the central
peak scales well with the magnetic field strength as the current in the solenoid is increased. 
Figure~\ref{lakeshorecalib} shows the peak frequency as a function of the magnetic field measured by an uncalibrated Hall probe
placed off axis within the solenoid bore. A linear fit to the data results in a root-mean-square deviation of only 1~MHz. 
We conclude that the position of the central peak (where the cyclotron heating is maximized) corresponds to the
cyclotron resonance frequency. We are able to measure this frequency to within 1~MHz in a uniform field, corresponding to a 
measurement of the magnetic field to 3.6 parts in $10^5$. Surprisingly, due to the nature of the observed lineshapes, we can 
identify the central peak frequency more precisely using hotter electron plasmas. 

\subsection{Neutral atom trap field}\label{neutraltrap}

One of the goals of the ALPHA collaboration is microwave spectroscopy of the hyperfine levels of antihydrogen's ground state. 
The highly inhomogeneous magnetic trapping fields, however, are detrimental for such a measurement. 
The inhomogeneity of the magnetic field and the strong field dependence of the hyperfine transition frequencies reduce
the effective time a trapped antihydrogen atom will be in resonance with a microwave field at a fixed frequency. 
In order to maximize the probability of inducing transitions between the hyperfine levels, a precise measurement of the magnetic 
trap minimum (where the field is most uniform) is critical. 

\begin{figure}
\begin{center}
\includegraphics{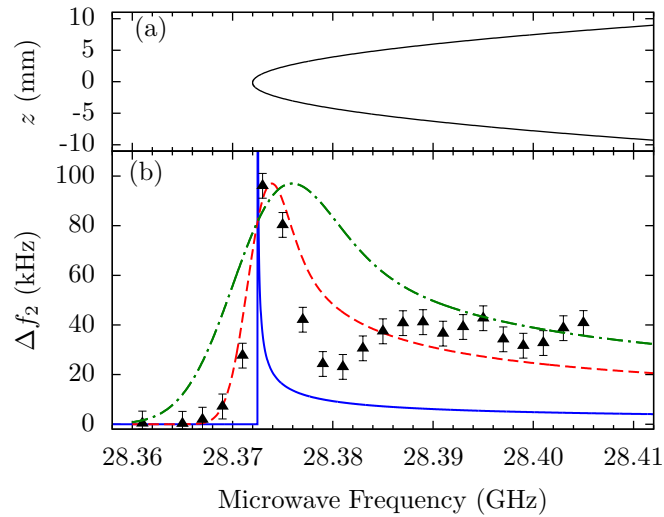}
\end{center}
\caption{(a) The axial positions of the cyclotron resonance as a function of the microwave frequency in the mirror field. 
(b) The measured cyclotron resonance lineshape (black triangles) and simple models of the expected lineshape 
in the inhomogeneous mirror coil magnetic field. The solid blue curve denotes the expected lineshape due to the magnetic field 
profile alone. The dashed and dot-dashed curves show the effect of thermal broadening of this lineshape with plasma temperatures 
of 25 K and 150 K, respectively. Here the microwave field has been taken to be a plane-wave propagating down the Penning trap axis.}
\label{MirLineshape}
\end{figure}

The neutral trap is formed by the superposition of an axial mirror field and an octupole field. Over the extent of 
the plasma, the octupole field varies by less than 0.1~mT so we first focus on the cyclotron frequency in the mirror 
field alone. The magnetic field produced by the mirror coils is given by 
\begin{equation} B_z(z,r) = B_0 + a\left(z^2 - \frac{r^2}{2}\right),\label{mirfield} \end{equation}
where $B_0$ is the magnetic field at $z=r=0$ and $a \approx 16$ $\mathrm{T/m^2}$ when the mirror coils 
are operated at the current used for antihydrogen trapping. 
The magnetic field is approximately uniform over the 1 mm plasma radius so the axial gradient of
the field will dominate. 
The magnetic field is most homogeneous at the minimum and microwaves tuned to this frequency will be resonant with the 
largest portion of the plasma. As the 
frequency is increased above the minimum, the microwaves come into resonance with increasingly narrow slices of plasma 
symmetrically displaced along the trap axis from the minimum. The axial position of the resonance is plotted as a function of 
cyclotron frequency in figure~\ref{MirLineshape}(a). 

A simple model of the expected lineshape can be constructed from the axial magnetic field profile with thermal broadening. 
The lineshape due to the magnetic field alone is shown in figure~\ref{MirLineshape}(b) (solid blue trace).
The thermal motion of the electrons parallel to the magnetic field will broaden this lineshape and introduce a 
systematic shift of the peak frequency away from the true minimum. This systematic shift arises
from the convolution of a Gaussian with the lineshape function due to the field profile. For example, if the microwave field is 
a plane wave propagating along the trap axis the FWHM of the Gaussian is given by the standard 
Doppler width $\Delta f_{\mathrm{FWHM}} = (8k_BT\ln2/mc^2)^{1/2}f_{\mathrm{c}}$, where $c$ is the speed of light in a vacuum. With a 
plasma temperature of 
150~K, this results in a shift of the peak frequency of 4~MHz above the true minimum resonance as illustrated in 
figure~\ref{MirLineshape}(b) (dot-dashed green trace). Without knowledge of the microwave mode structure, however, the true Doppler
width is unknown. The uniform field linewidths are narrower than predicted for the axially propagating plane wave 
case, suggesting that peak frequency is shifted by $<$ 4 MHz.

The observed cyclotron lineshape in the mirror coil field is also shown in figure~\ref{MirLineshape}(b). An onset peak is observed
as expected but the lineshape deviates from the simple model at higher frequencies. The distortion of the lineshape is a
result of spatial and frequency dependent variations in the CMEF. In section~\ref{Standingmap} the spatial variation of the CMEF is
measured and used to better model these lineshapes. 

\begin{figure}
\begin{center}
\includegraphics{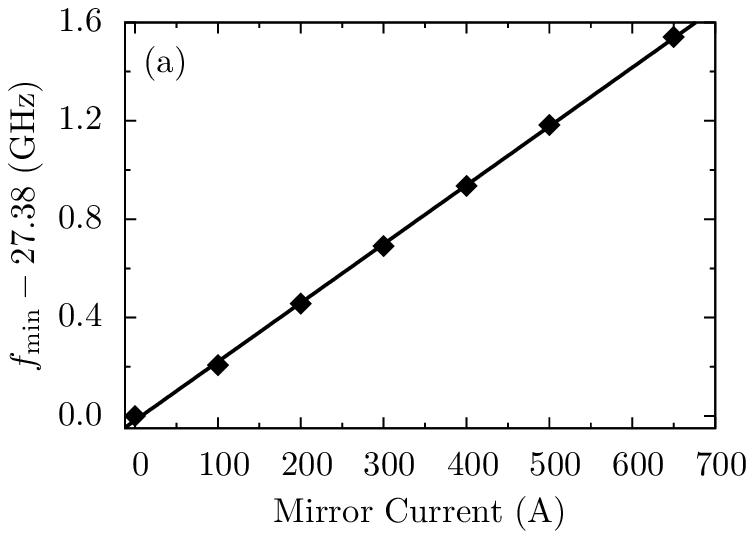}
\hspace{0.75cm}
\includegraphics{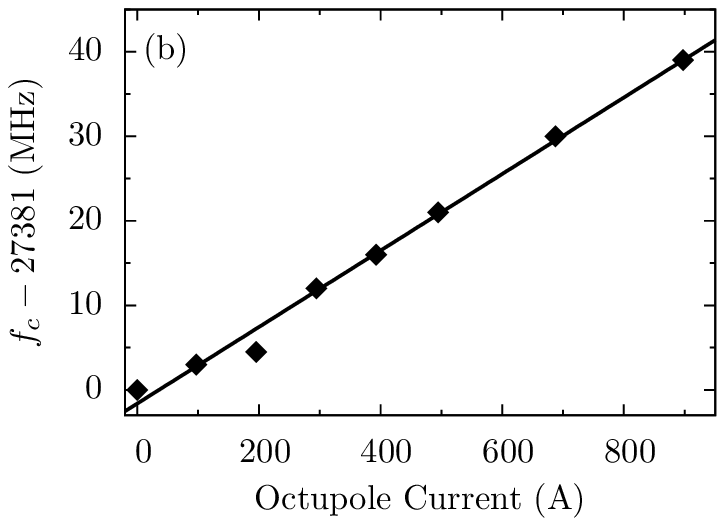}
\end{center}
\caption{(a) Frequency of the minimum cyclotron resonance ($f_{\mathrm{min}}$) as a function of the current in mirror coil magnets. 
The 
octupole magnet is not energized during these measurements. (b) The cyclotron resonance as a function of the current in the 
octupole magnet. The mirror coil magnets are not energized for these measurements.}
\label{cyc_vs_curr}
\end{figure}

While the full lineshape is significantly distorted, the low frequency onset we wish to characterize remains a prominent feature.
We take the position of the onset peak maximum to be the minimum cyclotron resonance frequency. This frequency is plotted
in figure~\ref{cyc_vs_curr}(a) as a function of the current in the mirror coils.
As the mirror current is increased, the minimum magnetic field increases and the field profile changes
from uniform across the plasma to highly non-uniform. Significant changes
in the local CMEF strengths will occur over the range of cyclotron frequencies plotted in Fig~\ref{cyc_vs_curr}(a). The onset peak 
frequency is relatively stable against these fluctuations, however, with a root-mean-squared deviation of 10~MHz obtained from
a linear fit to the data. Measurement of the mirror field lineshapes at plasma temperatures between 150 K and 1000 K show a 
broadening of the onset peak but we do not observe any systematic shift of the peak frequency. This is likely due to the strong 
effect of the changing CMEF amplitude as a function of position and frequency. We conclude that the rms deviation of 10MHz in 
figure~\ref{cyc_vs_curr}(a) reflects our uncertainty in determining the minimum cyclotron frequency. This corresponds to a 
relative magnetic field measurement of $\Delta B/B \approx 3.4\times10^{-4}$. 

We also characterized the contribution to the field by the octupole magnet. A perfect octupole field would have no  
axial component at the trap centre but the end turns in the octupole windings add a small amount. The  
octupole field is approximately uniform over the radius and length of the plasma so the lineshapes are effectively those of a
uniform field. Figure \ref{cyc_vs_curr}(b) shows the measured 
cyclotron frequency against the octupole current. At our nominal antihydrogen trapping current, the resonance is 
shifted by approximately 40~MHz. When the full neutral trap is energized one would expect that the minimum cyclotron resonance 
frequency 
would be a simple superposition of the octupole and mirror field resonances. Surprisingly, however, the minimum cyclotron resonance
is found roughly 40~MHz below the expected value. The cause of this deviation is currently unknown but may be due to some 
interaction between the superconducting magnets in the ALPHA apparatus such as shielding effects or flux pinning effects. 
While there are no known plasma effects that could explain this discrepancy, we cannot rule out the possibility of a systematic
offset of 40~MHz in the measurement of the minimum field in the full neutral trap.

\subsection{Cyclotron frequency shifts}\label{shifts}

So far we have been working under the assumption that the observed cyclotron frequency of the electron plasma is equivalent to 
the
single particle cyclotron frequency. In practice, however, a non-neutral plasma in a Penning trap can oscillate at a set 
of cyclotron modes that occur near the single particle frequency. These modes have been studied experimentally in 
electron~\cite{Gould1991} and magnesium ion plasmas~\cite{Sarid1995,Affolter2013} in a uniform magnetic field. 
The observed cyclotron modes have an angular dependence $\exp(i\ell\theta)$, where $\ell \ge 1$. 
Assuming a uniform density plasma out to a radius $r_{\mathrm{p}}$, these modes are shifted from the single particle cyclotron frequency
by an amount~\cite{Gould1994}
\begin{equation} \Delta f_{\mathrm{c},\ell} = \left[\ell - 1 - \left(\frac{r_{\mathrm{p}}}{r_{\mathrm{w}}}\right)^{2\ell}\right] f_{\mathrm{rot}}, \label{cycshift}
\end{equation}
where $f_{\mathrm{rot}}$ is the plasma rotation frequency and $r_{\mathrm{w}}$ is the inner radius of the electrodes. The $\ell = 1$ cyclotron 
mode is downshifted from the single particle cyclotron frequency by an amount equal to the diocotron frequency of the plasma: 
$\Delta f_{\mathrm{c},1} = -(r_{\mathrm{p}}/r_{\mathrm{w}})^2f_{\mathrm{rot}} = - f_{\mathrm{d}}$. Assuming a square plasma profile with a uniform density of 
$n = 9 \times 10^{13}\ \mathrm{m}^{-3}$ out to $r_{\mathrm{p}} = 1$ mm, we can estimate the rotation frequency 
as $f_{\mathrm{rot}} = nq/(4\pi\epsilon_0B) = 130$ kHz. Because the plasma radius is small compared to the electrode radius 
($r_{\mathrm{w}} = 22.5$ mm) 
the $\ell = 1$ cyclotron mode is downshifted by a negligible amount. The $\ell > 1$ modes, however, will be upshifted by 
integer multiples of 130 kHz, which is approaching the same order as the full spectral width of the 4 $\mu$s microwave pulses and 
the width of the observed central peaks (figure~\ref{uniformfield}(b)). We do not, however, observe any systematic shifts of the 
observed cyclotron lineshapes as a function of density between 
$n = 8\times10^{13}\ \mathrm{m}^{-3}$ and $n=2\times10^{14}\ \mathrm{m}^{-3}$. 

\section{Microwave electrometry}\label{fieldamp}

In addition to using the microwave-electron interactions to measure the cyclotron frequency, we can also extract information
about the microwave electric field. 
As previously discussed, the structure of a cylindrical Penning-Malmberg trap can give rise to large variations in microwave 
electric and magnetic field amplitudes as a function of frequency and position in the trap. For any microwave experiment in such 
an environment, including hyperfine spectroscopy of antihydrogen, \textit{in situ} diagnostics of the microwave field at different 
positions and frequencies can be extremely useful. 

\subsection{Electric field amplitude}

Using the quadrupole mode frequency calibration, we can measure the change in temperature due to a microwave
pulse and infer the CMEF amplitude from (\ref{short_soln}) and (\ref{DT}). To ensure we are in the short pulse limit, we 
inject 80 ns microwave pulses. The collisional rate at which the cyclotron motion of the electrons equilibrates with the 
motion parallel to the 1 T field at 150 K is approximately~\cite{Glinsky1992} 
$\Gamma \sim 10^{-9}n \ \mathrm{m^3s^{-1}}$. We use a plasma with a 
density of $n =  2\times10^{14} \ \mathrm{m}^{-3}$ giving an expected equilibration rate of $\Gamma \sim 2\times10^{5} \ \mathrm{s^{-1}}$. 
The Agilent 8257D 
synthesizer produces a stable frequency and phase over the duration of the microwave pulse so the spectral width is effectively 
set by the pulse length. 
The full spectral width of the 80 ns pulse is $2.5\times10^7$ Hz, two orders of magnitude larger than $\Gamma$, so collisional
damping can be neglected and (\ref{short_soln}) is valid. 

We inject square microwave pulses where the transverse components of the electric field approximately take the form: 
\begin{eqnarray} E_x(t) &= E_{x,0}\cos(\omega_0t)[H(t+\tau/2) - H(t - \tau/2)], \\
              E_y(t) &= E_{y,0}\cos(\omega_0t+\delta_y)[H(t+\tau/2) - H(t - \tau/2)],
\end{eqnarray}
where $H$ is the Heaviside step function and $\tau$ is the pulse duration. The co-rotating component of the
electric field is given by $E_+(t) = E_x(t) + iE_y(t)$ and one finds

\begin{equation}
\int_{-\infty}^{\infty}E_+(t')e^{i\omega t'}dt' = \left(\frac{\sin[(\omega_0-\omega)\tau/2]}{
\omega_0 - \omega}  + \frac{\sin[(\omega_0+\omega)\tau/2]}{\omega_0+\omega}\right)E_0,
\end{equation}
where $E_0 = E_{x,0} +iE_{y,0}e^{i\delta_y}$.
Near resonance $\Delta \omega = \omega_0 - \omega \ll \omega_0 + \omega$ so to good approximation
\begin{equation} \int_{-\infty}^{\infty}E_+(t')e^{i\omega t'}dt' = \frac{\tau}{2}\mathrm{sinc}\left(\frac{\Delta \omega \tau}{2}\right)E_0
\label{E+}, \end{equation}
with a total error of order $1/(\omega_0 \tau)$.
Inserting (\ref{E+}) into (\ref{short_soln}) we have
\begin{equation} \Delta \langle \mathrm{KE_{\perp}} \rangle = \frac{q^2\tau^2}{8m}
\mathrm{sinc}^2\left(\frac{\Delta \omega \tau}{2}\right)|E_0|^2.
\label{DKE}\end{equation}
Using (\ref{DW}), (\ref{DT}) and (\ref{DKE}) and solving for the amplitude of the
CMEF at resonance ($\Delta \omega = 0$) yields
\begin{equation} |E_0| = \frac{2\sqrt{3mk_B\Delta f_2/\beta}}{q\tau} \label{efield}.
\end{equation}

As an example we use a plasma of $1.2\times10^7$ electrons with a measured quadrupole frequency 
calibration of $\beta^{-1} = 3.7$ K/kHz. Microwaves are injected at a resonant frequency of 
27.370 GHz in $80$ ns pulses. With a power of $9$ mW emitted from the microwave horn, using 
the low-power microwave transmission path, we measure a quadrupole mode
shift of $\Delta f_2 = 100$ kHz, corresponding to a CMEF amplitude of $18.4$~Vm$^{-1}$. Approximating the microwaves as plane waves 
propagating down the trap axis, this corresponds to a power of $0.7$ mW; a loss of $11$ dB from the horn to the plasma. As the
resonance frequency changes, large fluctuations in electric field amplitude are observed. 

For hyperfine spectroscopy of trapped antihydrogen it is useful to estimate the hyperfine transition rate expected. 
In this case, the transverse magnetic field component of the microwave field is the relevant quantity. Unfortunately,
without knowledge of the field structure in the trap we cannot properly calculate the magnetic field amplitude from the measured
electric field. We can make an order of magnitude estimate, however, by approximating the microwaves as plane-waves in free space. 
Based on the measured CMEF amplitude, the high power transmission path of the microwave system 
would produce a CMEF of roughly 100 Vm$^{-1}$. For a plane wave 
with this electric field, the co-rotating component of the magnetic field is $B = E/c \approx 0.33$~$\mu$T, where
$c$ is the speed of light in a vacuum. 
Based on a simulation of the interaction of microwaves with trapped antihydrogen, a positron spin flip rate of approximately 
1 s$^{-1}$ is expected, consistent with the observations 
in~\cite{Amole2012}. While only an order of magnitude estimate, these measurements are very useful in a situation where 
the microwave field is otherwise unknown.

\subsection{Microwave electric field maps}\label{Standingmap}

By moving the plasma along the trap axis and repeating the electric field amplitude measurement described above, the axial 
dependence of the CMEF can be probed.
The spatial resolution will be set by the plasma length as field variations on a smaller scale will be averaged out. For the
typical plasma used in the current work, we can therefore sample changes in the CMEF over 2 - 4 cm in this manner. 

\begin{figure}
\centering \includegraphics{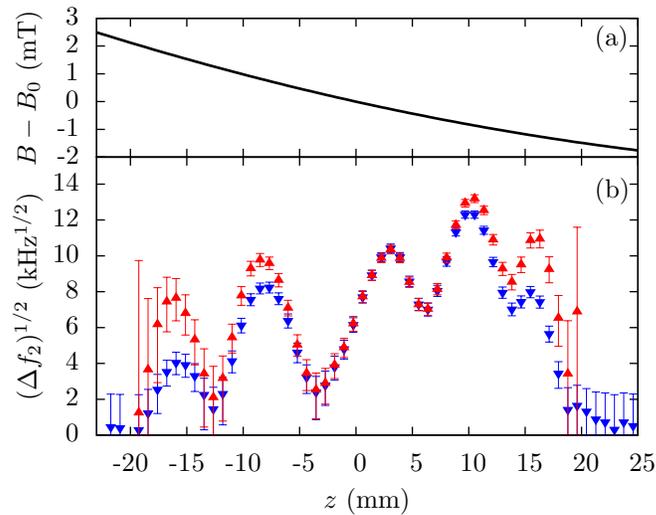}
\caption{(a) Magnetic field gradient across the electron plasma created by the fringe field of 
the catching solenoid (see figure~\ref{setup}). (b) The square root of the measured quadrupole mode frequency
increase plotted as a function of axial position over the plasma for a microwave frequency of 28.375~GHz. 
Assuming a perfect linear magnetic field gradient and a cylindrical plasma, the measured $(\Delta f_2)^{1/2}$ (inverted blue
triangles) are proportional to the CMEF amplitude. The red triangles are the relative CMEF amplitudes when the spheroidal
plasma shape and deviations from a linear gradient are considered.}
\label{StandingWaveMap}
\end{figure}

We can probe the electric field on a finer scale by keeping the plasma position fixed and applying a magnetic field gradient 
across the plasma (see figure~\ref{StandingWaveMap}(a)) such that only a small portion 
of the plasma is resonant at a given frequency. 
The gradient is produced by the fringe field of a superconducting solenoid at one end of the ALPHA trap 
(see figure~\ref{setup}) and is numerically modelled using TOSCA/OPERA3D~\cite{OPERA}. 
Microwaves are pulsed every 35 seconds at a fixed frequency, while the Penning trap solenoid is slowly swept (keeping the 
gradient fixed) to bring different parts of the plasma into resonance for each pulse. The resonant position of each pulse
is determined by the modelled magnetic field gradient and the rate at which the solenoid is swept.
A microwave pulse length of 4 $\mu$s with a full spectral width of 500 kHz is employed such that only a small 
portion of the plasma is excited with each pulse. With a uniform microwave electric field, this scan 
would be analogous to magnetic resonance imaging of the plasma. Here, however, the plasma is approximately a
uniform density spheroid in a highly structured electric field. By scanning the resonance across the plasma and measuring 
the quadrupole frequency shifts, a relative map of the CMEF along the $z$-axis of the plasma can be generated. As a simple
estimate we can assume a perfectly linear magnetic field gradient and a cylindrical plasma (uniform radius between $z=-L/2$ and 
$z=L/2$) such that the relative plasma heating only depends on the local CMEF amplitude. 

In reality, the plasma is better approximated by a spheroid and the fringe field of the solenoid doesn't produce a perfectly linear 
magnetic field gradient. 
These two effects change the volume of the plasma that is in resonance with a microwave pulse as a function of $z$, 
increasing or decreasing the observed quadrupole frequency shift. To account for the spheroidal shape of the plasma, we multiply by 
a correction factor $r(0)/r(z) = (1-(2z/L)^2)^{-1/2}$, where $L = 40$ mm for the plasma used in figure~\ref{StandingWaveMap}. 
As the slope of the fringe magnetic field reduces at
high $z$, a larger portion of the plasma will be excited by each pulse, increasing the plasma heating. 
This can be corrected for by multiplying the observed response by a factor $(B'(z)/B'(0))^{1/2}$, where $B'(z) = dB/dz$. Both
correction factors have been normalized to the response at the centre of the plasma. 
As an example, figure~\ref{StandingWaveMap}(b) plots $(\Delta f_2)^{1/2}$ (which is proportional to $|E_0|$) as a function of $z$ at 
a microwave frequency of 28.375 GHz with and without the corrections applied. 
The spheroidal shape correction has the greatest effect far from the middle of the plasma and breaks down when $|z| = L/2$. Better 
measurement of the CMEF strength at these points can be obtained by repositioning the plasma over the region of interest. 

The spatial resolution of this mapping is set by the field gradient and the linewidth of the resonance. 
In the current example, the gradient used is approximately 0.09 mT/mm. Based on the FWHM of the observed 
uniform field lineshape at 140 K (see figure~\ref{uniformfield}), which is 0.2 mT in terms of magnetic field, we estimate that
each pulse samples a slice of plasma approximately 2 mm long. 

\subsection{Modelling the mirror field lineshapes}

With the measurements of the axial CMEF profile we can attempt to better model the cyclotron lineshapes observed in 
section~\ref{neutraltrap}. Starting with the simple lineshape model discussed in section~\ref{neutraltrap} we apply a frequency 
dependant correction factor based on a measured CMEF map. In the mirror field, each microwave frequency is resonant
with two slices of the plasma symmetric about $z = 0$ (see figure~\ref{MirLineshape}(a)). From the CMEF map we can estimate 
the relative field strengths at these two positions and therefore the distortion of the lineshape due to the spatially varying 
electric field. This model will not be completely accurate, however, as the CMEF is mapped at a fixed microwave frequency. As the
frequency is changed during the cyclotron resonance lineshape measurement the CMEF profile will change with it. Thermal broadening 
is included in the model by convolving a Gaussian with the lineshape based on the magnetic field profile alone. Because we do not 
have enough information about the structure of the microwave field to accurately model the thermal broadening, a generic Gaussian 
given by $\exp{(-4\ln2f^2/{\Delta f}^2)}$ is used, where the width $\Delta f$ is a fit parameter.

\begin{figure}
\includegraphics[width=0.48\columnwidth]{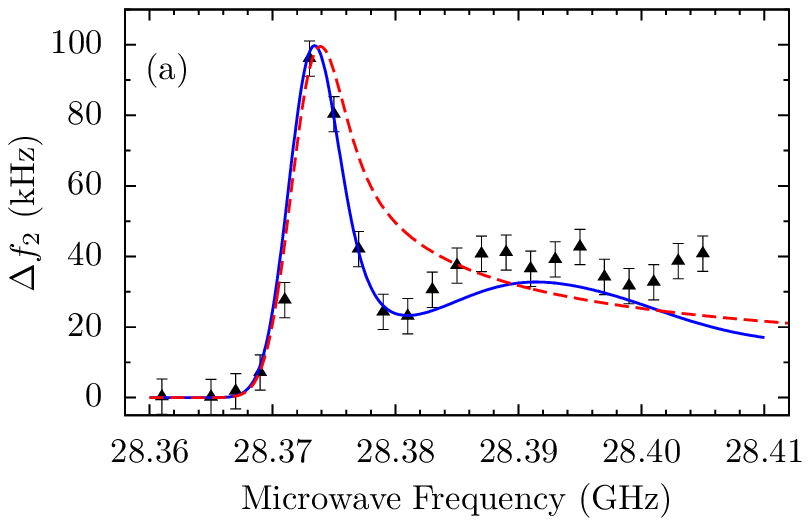}\hspace{0.04\columnwidth}\includegraphics[width=0.48\columnwidth]{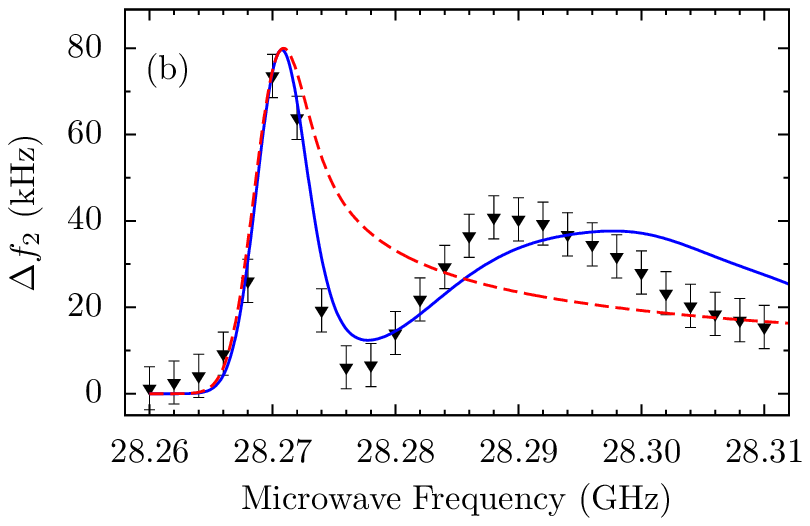}
\caption{(a) The improved model (blue line) of the measured cyclotron lineshape shown in figure~\ref{MirLineshape} 
using a map of the CMEF at 28.375 GHz. The red dashed line shows the simple model for comparison. (b) A second example 
demonstrating the improved modelling (blue line) of the cyclotron resonance lineshape using a CMEF map at 28.270 GHz compared
to the simple model (red dashed line).}
\label{ModelFits}
\end{figure}

Using the CMEF map at 28.375 GHz shown in figure~\ref{StandingWaveMap}, a model for the cyclotron lineshape shown in 
figure~\ref{MirLineshape} (with a peak response frequency at 28.372 GHz) is generated (figure~\ref{ModelFits}(a)). The model 
matches the onset peak structure well and is an improvement over the simple model but still deviates from the measurements above a 
frequency of 28.380 GHz. This is likely the effect of the 
CMEF profile changing as a function of frequency. Figure~\ref{ModelFits}(b) shows a second example of a modelled lineshape with a 
peak response frequency at 28.270 GHz and using a CMEF map at 28.270 GHz. The improved agreement with the measured cyclotron 
resonance lineshapes provides an additional measure of confidence that the observed lineshapes are due to the magnetic field
inhomogeneity, thermal broadening, and the spatial and frequency dependence of the CMEF amplitude. 

\section{Conclusion}

While the uniform field lineshapes measured in section~\ref{Uniform} are not fully understood, we can identify the cyclotron 
resonance frequency to within 1 MHz. This is of the same order as the spectral width of the $4$ $\mu$s microwave
pulses and may be improved with longer pulses. This is also approaching the order on which systematic shifts of the
observed resonance away from the single particle cyclotron frequency, due to the plasma rotation, are expected. If
the resolution of the cyclotron frequency measurement is increased further, careful study of the frequency shifts
will be necessary. 

When the magnetic field is non-uniform over the plasma length, the spatial dependence of the microwave electric field will distort 
the cyclotron lineshape significantly. In the neutral atom trap field, the uncertainty in the measurement of the minimum field due 
to these distortions can be reduced by flattening the field at the 
minimum. With
more of the plasma resonant at the minimum cyclotron frequency, variations in CMEF strength will be averaged over a larger
range, approaching the uniform field case, and make identification of the onset peak easier. A new version of the
ALPHA apparatus, currently under construction, includes three additional mirror coils that can act as compensation coils to 
flatten the field minimum while maintaining the magnetic trap depth. 

Eliminating the uncertainties resulting from the spatial dependence of the microwave field requires the 
inclusion of a microwave cavity. In addition to removing a large source of uncertainty, 
if the majority of the plasma is confined between nodes of a trapped resonator mode, the lineshape 
will be dominated by a Doppler free peak at the cyclotron frequency~\cite{Kleppner1962}, greatly increasing the achievable 
resolution of the measurement. Designing a cavity that does not compromise the ability to store and manipulate charged plasmas 
presents a challenge but may be included in future upgrades to the ALPHA apparatus. 

In this work, we have described a novel method for the measurement of the cyclotron frequency of an electron plasma in a 
Penning-Malmberg trap. This method is applied at microwave cyclotron frequencies as an \textit{in situ}, non-destructive, and spatially
resolving measurement of the static magnetic field and microwave electric field strengths in the trap. In the ALPHA trap our
measurement of the magnetic field had an accuracy of about 3.6 parts in 10$^5$ for the nominally uniform magnetic field.
In the magnetic neutral atom trap fields, the minimum was resolved to within about 3.4 parts in $10^4$, with a potential systematic
offset of 1.4 parts in $10^3$ which cannot be ruled out at this time.  An uncertainty of 3.4 parts in $10^4$ would translate to an 
inaccuracy of only 64 Hz (2.5 parts in $10^{14}$) in the 1S - 2S transition frequency, assuming a minimum magnetic field 
of 1 T~\cite{Cesar2001}. With hardware improvements and further study, the sensitivity of these measurements could approach a 
resolution limited by collisional scattering (roughly 1 part in $10^6$ for typical plasmas used here). Implementation of these 
techniques as a diagnostic tool requires an electron or ion plasma with a detectable quadrupole mode frequency and a method for 
excitation of the cyclotron motion.
 
\section{Acknowledgements}

This work was supported by CNPq, FINEP/RENAFAE (Brazil), ISF (Israel), 
FNU (Denmark), VR (Sweden), NSERC, NRC/TRIUMF, AITF, FQRNT
(Canada), DOE, NSF (USA), EPSRC, the Royal Society and the 
Leverhulme Trust (UK). 

\section{References}

\bibliographystyle{unsrt.bst}
\bibliography{ecr_refs}

\begin{thebibliography}{10}

\bibitem{Marshall1998}
A.~G. Marshall, C.~L. Hendrickson, and G.~S. Jackson.
\newblock Fourier transform ion cyclotron resonance mass spectrometry: A
  primer.
\newblock {\em Mass Spectrom. Rev.}, 17(1):1--35, 1998.

\bibitem{Bergstrom2002}
I.~Bergstr{\"o}m, C.~Carlberg, T.~Fritioff, G.~Douysset, J.~Sch{\"o}nfelder,
  and R.~Schuch.
\newblock {SMILETRAP} -- {A} {P}enning trap facility for precision mass
  measurements using highly charged ions.
\newblock {\em Nucl. Instrum. Meth. A}, 487(3):618--651, 2002.

\bibitem{Dilling2006}
J.~Dilling, R.~Baartman, P.~Bricault, M.~Brodeur, L.~Blomeley, F.~Buchinger,
  J.~Crawford, JR~Crespo L{\'o}pez-Urrutia, P.~Delheij, M.~Froese, et~al.
\newblock Mass measurements on highly charged radioactive ions, a new approach
  to high precision with {TITAN}.
\newblock {\em Int. J. Mass Spectrom.}, 251(2):198--203, 2006.

\bibitem{DiSciacca2012}
J.~DiSciacca and G.~Gabrielse.
\newblock Direct measurement of the proton magnetic moment.
\newblock {\em Phys. Rev. Lett.}, 108:153001, 2012.

\bibitem{Odom2006}
B.~Odom, D.~Hanneke, B.~D'Urso, and G.~Gabrielse.
\newblock New measurement of the electron magnetic moment using a one-electron
  quantum cyclotron.
\newblock {\em Phys. Rev. Lett.}, 97:030801, 2006.

\bibitem{Gould1991}
R.~W. Gould and M.~A. LaPointe.
\newblock Cyclotron resonance in a pure electron plasma column.
\newblock {\em Phys. Rev. Lett.}, 67:3685--3688, 1991.

\bibitem{Sarid1995}
E.~Sarid, F.~Anderegg, and C.~F. Driscoll.
\newblock Cyclotron resonance phenomena in a non-neutral multispecies ion
  plasma.
\newblock {\em Phys. Plasmas}, 2(8):2895--2907, 1995.

\bibitem{Dyck1981}
R.~S. Van~Dyck and P.~B. Schwinberg.
\newblock Preliminary proton/electron mass ratio using a compensated quadring
  penning trap.
\newblock {\em Phys. Rev. Lett.}, 47:395--398, 1981.

\bibitem{Friesen2013}
T.~Friesen, C.~Amole, M.~D. Ashkezari, M.~Baquero-Ruiz, W.~Bertsche, P.~D.
  Bowe, E.~Butler, A.~Capra, C.~L. Cesar, M.~Charlton, A.~Deller, N.~Evetts,
  S.~Eriksson, J.~Fajans, M.~C. Fujiwara, D.~R. Gill, A.~Gutierrez, J.~S.
  Hangst, W.~N. Hardy, M.~E. Hayden, C.~A. Isaac, S.~Jonsell, L.~Kurchaninov,
  A.~Little, N.~Madsen, J.~T.~K. McKenna, S.~Menary, S.~C. Napoli,
  K.~Olchanski, A.~Olin, P.~Pusa, C.~{\O}. Rasmussen, F.~Robicheaux, E.~Sarid,
  D.~M. Silveira, C.~So, S.~Stracka, R.~I. Thompson, D.~P. van~der Werf, and
  J.~S. Wurtele.
\newblock Electron plasmas as a diagnostic tool for hyperfine spectroscopy of
  antihydrogen.
\newblock {\em AIP Conference Proceedings}, 1521(1):123--133, 2013.

\bibitem{Pritchard1983}
D.E. Pritchard.
\newblock Cooling neutral atoms in a magnetic trap for precision spectroscopy.
\newblock {\em Phys. Rev. Lett.}, 51:1336--1339, 1983.

\bibitem{Andresen2010}
G.~B. Andresen, M.~D. Ashkezari, M.~Baquero-Ruiz, W.~Bertsche, P.~D. Bowe,
  E.~Butler, C.~L. Cesar, S.~Chapman, M.~Charlton, A.~Deller, S.~Eriksson,
  J.~Fajans, T.~Friesen, M.~C. Fujiwara, D.~R. Gill, A.~Gutierrez, J.~S.
  Hangst, W.~N. Hardy, M.~E. Hayden, A.~J. Humphries, R.~Hydomako, M.~J.
  Jenkins, S.~Jonsell, L.~V. J\o{}rgensen, L.~Kurchaninov, N.~Madsen,
  S.~Menary, P.~Nolan, K.~Olchanski, A.~Olin, A.~Povilus, P.~Pusa,
  F.~Robicheaux, E.~Sarid, S.~{Seif El Nasr}, D.~M. Silveira, C.~So, J.~W.
  Storey, R.~I. Thompson, D.~P. van~der Werf, J.~S. Wurtele, and Y.~Yamazaki.
\newblock Trapped antihydrogen.
\newblock {\em Nature}, 468(7324):673--676, 2010.

\bibitem{Gabrielse2012}
G.~Gabrielse, R.~Kalra, W.~S. Kolthammer, R.~McConnell, P.~Richerme,
  D.~Grzonka, W.~Oelert, T.~Sefzick, M.~Zielinski, D.~W. Fitzakerley, M.~C.
  George, E.~A. Hessels, C.~H. Storry, M.~Weel, A.~M\"ullers, and J.~Walz.
\newblock Trapped antihydrogen in its ground state.
\newblock {\em Phys. Rev. Lett.}, 108:113002, 2012.

\bibitem{Ashkezari2012}
M.~D. Ashkezari, G.~B. Andresen, M.~Baquero-Ruiz, W.~Bertsche, P.~D. Bowe,
  E.~Butler, C.~L. Cesar, S.~Chapman, M.~Charlton, A.~Deller, S.~Eriksson,
  J.~Fajans, T.~Friesen, M.C. Fujiwara, D.R. Gill, A.~Gutierrez, J.S. Hangst,
  W.N. Hardy, R.S. Hayano, M.E. Hayden, A.J. Humphries, R.~Hydomako,
  S.~Jonsell, L.~Kurchaninov, N.~Madsen, S.~Menary, P.~Nolan, K.~Olchanski,
  A.~Olin, A.~Povilus, P.~Pusa, F.~Robicheaux, E.~Sarid, D.M. Silveira, C.~So,
  J.W. Storey, R.I. Thompson, D.P. van~der Werf, J.S. Wurtele, and Y.~Yamazaki.
\newblock Progress towards microwave spectroscopy of trapped antihydrogen.
\newblock {\em Hyperfine Interactions}, 212(1-3):81--90, 2012.

\bibitem{Amole2012}
C.~Amole, M.~D. Ashkezari, M.~Baquero-Ruiz, W.~Bertsche, P.~D. Bowe, E.~Butler,
  A.~Capra, C.~L. Cesar, M.~Charlton, A.~Deller, P.~H. Donnan, S.~Eriksson,
  J.~Fajans, T.~Friesen, M.~C. Fujiwara, D.~R. Gill, A.~Gutierrez, J.~S.
  Hangst, W.~N. Hardy, M.~E. Hayden, A.~J. Humphries, C.~A. Isaac, S.~Jonsell,
  L.~Kurchaninov, A.~Little, N.~Madsen, J.~T.~K. McKenna, S.~Menary, S.~C.
  Napoli, P.~Nolan, K.~Olchanski, A.~Olin, P.~Pusa, C.~{\O}. Rasmussen,
  F.~Robicheaux, E.~Sarid, C.~R. Shields, D.~M. Silveira, S.~Stracka, C.~So,
  R.~I. Thompson, D.~P. van~der Werf, and J.~S. Wurtele.
\newblock Resonant quantum transitions in trapped antihydrogen atoms.
\newblock {\em Nature}, 483(7390):439--443, 2012.

\bibitem{Dubin1991}
D.~H.~E. Dubin.
\newblock Theory of electrostatic fluid modes in a cold spheroidal non-neutral
  plasma.
\newblock {\em Phys. Rev. Lett.}, 66(16):2076--2079, 1991.

\bibitem{Tinkle1994}
M.~D. Tinkle, R.~G. Greaves, C.~M. Surko, R.~L. Spencer, and G.~W. Mason.
\newblock Low-order modes as diagnostics of spheroidal non-neutral plasmas.
\newblock {\em Phys. Rev. Lett.}, 72:352--355, 1994.

\bibitem{Amoretti2003}
M.~Amoretti, C.~Amsler, G.~Bonomi, A.~Bouchta, P.~D. Bowe, C.~Carraro, C.~L.
  Cesar, M.~Charlton, M.~Doser, V.~Filippini, A.~Fontana, M.~C. Fujiwara,
  R.~Funakoshi, P.~Genova, J.~S. Hangst, R.~S. Hayano, L.~V. J\o{}rgensen,
  V.~Lagomarsino, R.~Landua, D.~Lindel\"of, E.~Lodi Rizzini, M.~Macr\'\i{},
  N.~Madsen, G.~Manuzio, P.~Montagna, H.~Pruys, C.~Regenfus, A.~Rotondi,
  G.~Testera, A.~Variola, and D.~P. van~der Werf.
\newblock Positron plasma diagnostics and temperature control for antihydrogen
  production.
\newblock {\em Phys. Rev. Lett.}, 91(5):055001, 2003.

\bibitem{Speck2007}
A.~Speck, G.~Gabrielse, P.~Larochelle, D.~Le Sage, B.~Levitt, W.S. Kolthammer,
  R.~McConnell, J.~Wrubel, D.~Grzonka, W.~Oelert, T.~Sefzick, Z.~Zhang,
  D.~Comeau, M.C. George, E.A. Hessels, C.H. Storry, M.~Weel, and J.~Walz.
\newblock Density and geometry of single component plasmas.
\newblock {\em Phys. Lett. B}, 650(2-3):119, 2007.

\bibitem{Dubin1993}
D.~H.~E. Dubin.
\newblock Equilibrium and dynamics of uniform density ellipsoidal non-neutral
  plasmas.
\newblock {\em Phys. Fluids B}, 5(2):295--324, 1993.

\bibitem{Tinkle1995}
M.~D. Tinkle, R.~G. Greaves, and C.~M. Surko.
\newblock Low-order longitudinal modes of single-component plasmas.
\newblock {\em Phys. of Plasmas}, 2(8):2880--2894, 1995.

\bibitem{Amole2014}
C.~Amole, G.B. Andresen, M.D. Ashkezari, M.~Baquero-Ruiz, W.~Bertsche, P.D.
  Bowe, E.~Butler, A.~Capra, P.T. Carpenter, C.L. Cesar, S.~Chapman,
  M.~Charlton, A.~Deller, S.~Eriksson, J.~Escallier, J.~Fajans, T.~Friesen,
  M.C. Fujiwara, D.R. Gill, A.~Gutierrez, J.S. Hangst, W.N. Hardy, R.S. Hayano,
  M.E. Hayden, A.J. Humphries, J.L. Hurt, R.~Hydomako, C.A. Isaac, M.J.
  Jenkins, S.~Jonsell, L.V. Jørgensen, S.J. Kerrigan, L.~Kurchaninov,
  N.~Madsen, A.~Marone, J.T.K. McKenna, S.~Menary, P.~Nolan, K.~Olchanski,
  A.~Olin, B.~Parker, A.~Povilus, P.~Pusa, F.~Robicheaux, E.~Sarid, D.~Seddon,
  S.~Seif~El Nasr, D.M. Silveira, C.~So, J.W. Storey, R.I. Thompson,
  J.~Thornhill, D.~Wells, D.P. van~der Werf, J.S. Wurtele, and Y.~Yamazaki.
\newblock The {ALPHA} antihydrogen trapping apparatus.
\newblock {\em Nucl. Instrum. Meth. A}, 735(0):319 -- 340, 2014.

\bibitem{Bertsche2006}
W.~Bertsche, A.~Boston, P.D. Bowe, C.L. Cesar, S.~Chapman, M.~Charlton,
  M.~Chartier, A.~Deutsch, J.~Fajans, M.C. Fujiwara, R.~Funakoshi,
  K.~Gomberoff, J.S. Hangst, R.S. Hayano, M.J. Jenkins, L.V. J\o{}rgensen,
  P.~Ko, N.~Madsen, P.~Nolan, R.D. Page, L.G.C. Posada, A.~Povilus, E.~Sarid,
  D.M. Silveira, D.P. van~der Werf, Y.~Yamazaki, B.~Parker, J.~Escallier, and
  A.~Ghosh.
\newblock A magnetic trap for antihydrogen confinement.
\newblock {\em Nucl. Inst. Meth. A}, 566(2):746 -- 756, 2006.

\bibitem{Andresen2009}
G.~B. Andresen, W.~Bertsche, P.~D. Bowe, C.~C. Bray, E.~Butler, C.~L. Cesar,
  S.~Chapman, M.~Charlton, J.~Fajans, M.~C. Fujiwara, D.~R. Gill, J.~S. Hangst,
  W.~N. Hardy, R.~S. Hayano, M.~E. Hayden, A.~J. Humphries, R.~Hydomako, L.~V.
  J\o{}rgensen, S.~J. Kerrigan, L.~Kurchaninov, R.~Lambo, N.~Madsen, P.~Nolan,
  K.~Olchanski, A.~Olin, A.~Povilus, P.~Pusa, E.~Sarid, S.~{Seif El Nasr},
  D.~M. Silveira, J.~W. Storey, R.~I. Thompson, D.~P. van~der Werf, and
  Y.~Yamazaki.
\newblock Antiproton, positron, and electron imaging with a microchannel plate
  phosphor detector.
\newblock {\em Rev. Sci. Instrum.}, 80(12):123701, 2009.

\bibitem{Prasad1979}
S.~A. Prasad and T.~M. O'Neil.
\newblock Finite length thermal equilibria of a pure electron plasma column.
\newblock {\em Phys. Fluids}, 22(2):278--281, 1979.

\bibitem{Huang1997}
X.-P. Huang, F.~Anderegg, E.~M. Hollmann, C.~F. Driscoll, and T.~M. O'Neil.
\newblock Steady-state confinement of non-neutral plasmas by rotating electric
  fields.
\newblock {\em Phys. Rev. Lett.}, 78:875--878, 1997.

\bibitem{Eggleston1992}
D.~L. Eggleston, C.~F. Driscoll, B.~R. Beck, A.~W. Hyatt, and J.~H. Malmberg.
\newblock Parallel energy analyzer for pure electron plasma devices.
\newblock {\em Phys. of Fluids B}, 4(10):3432--3439, 1992.

\bibitem{Glinsky1992}
M.~E. Glinsky, T.~M. O'Neil, M.~N. Rosenbluth, K.~Tsuruta, and S.~Ichimaru.
\newblock Collisional equipartition rate for a magnetized pure electron plasma.
\newblock {\em Phys. Fluids B: Plasma Phys.}, 4(5):1156--1166, 1992.

\bibitem{Kleppner1962}
D.~Kleppner, H.~M. Goldenberg, and N.F. Ramsey.
\newblock Theory of the hydrogen maser.
\newblock {\em Phys. Rev.}, 126:603--615, 1962.

\bibitem{Dicke1953}
R.~H. Dicke.
\newblock The effect of collisions upon the {Doppler} width of spectral lines.
\newblock {\em Phys. Rev.}, 89:472--473, 1953.

\bibitem{Affolter2013}
M.~Affolter, F.~Anderegg, C.~F. Driscoll, and D.~H.~E. Dubin.
\newblock Cyclotron resonances in a non-neutral multispecies ion plasma.
\newblock {\em AIP Conference Proceedings}, 1521(1):175--183, 2013.

\bibitem{Gould1994}
R.~W. Gould.
\newblock Theory of cyclotron resonance in a cylindrical non-neutral plasma.
\newblock {\em Phys. Plasmas}, 2(5):1404--1411, 1995.

\bibitem{OPERA}
Commercial product from Cobham Technical Services,
  http://www.cobham.com/technicalservices.

\bibitem{Cesar2001}
C.~L. Cesar.
\newblock Zeeman effect on the $1s-2s$ transition in trapped hydrogen and
  antihydrogen.
\newblock {\em Phys. Rev. A}, 64:023418, 2001.

\end{thebibliography}

\end{document}